\documentclass[12pt]{article}
\usepackage{times}
\usepackage{graphicx}
\usepackage{array, ltablex, multirow}
\usepackage[numbers,square,sort&compress]{natbib}
\usepackage{hyperref}
\usepackage{float}
\usepackage{authblk}
\usepackage{xcolor}
\usepackage{siunitx}
\DeclareSIUnit{\electrons}{e\textsuperscript{$-$}}

\date{}

\begin{document}
\title{Direct visualization of the 3D structure of silicon impurities in graphene}

\author{Christoph Hofer}
\author{Viera Skakalova}
\author{Mohammad Reza Ahmadpour Monazam}
\author{Clemens Mangler}
\author{Jani Kotakoski}
\author{Toma Susi}
\author{Jannik C. Meyer}

\affil[1]{Faculty of Physics, University of Vienna, Boltzmanngasse 5, A-1090, Vienna, Austria}

\maketitle

\begin{abstract}
We directly visualize the three-dimensional (3D) geometry and dynamics of silicon impurities in graphene as well as their dynamics by aberration-corrected scanning transmission electron microscopy. By acquiring images when the sample is tilted, we show that an asymmetry of the atomic position of the heteroatom in the projection reveals the non-planarity of the structure. From a sequence of images, we further demonstrate that the Si atom switches between up- and down- configurations with respect to the graphene plane, with an asymmetric cross-section. We further analyze the 3D structure and dynamics of a silicon tetramer in graphene.  Our results clarify the out-of-plane structure of impurities in graphene by direct experimental observation and open a new route to study their dynamics in three dimensions.
\end{abstract}

\section{Introduction}

Although the extraordinary properties of pristine graphene have raised enormous interest in the scientific community and industry, most applications require modified properties such as a non-zero bandgap \cite{Zhou2007,Schwierz2010}. One approach to tailor graphene properties is doping with heteroatoms, which may open a bandgap  \cite{Azadeh2011,Shahrokhi2017,Houmad2015,Zhang2016} or enhance local plasmon resonances \cite{Grigorenko2012}. To understand the influence of atomic substitutions, the three-dimensional (3D) position of every atom has to be determined.  The 3D structure as well as the beam-induced dynamics are also important in the context of single atom manipulation in graphene \cite{Susi2014,Susi2017,Tripathi2018a}.
 
Out-of-plane buckling of heteroatoms has already been demonstrated indirectly by analyzing the finestructure of electron energy loss spectra \cite{Ramasse2013,Tripathi2018,Susi2017a,Zhou2012}. Nevertheless, to fully understand the real magnitude of the displacement from the graphene plane as well as the local lattice distortion, there is a need for directly measuring the 3D structure. Although recent advances in scanning transmission electron microscopy (STEM) allow extracting structural and chemical information at the atomic scale \cite{Krivanek1999,Colliex2014}, this is only possible as a two-dimensional projection of the object. Electron tomography allows to reconstruct the 3D atomic structure \cite{Bals2014,Miao2016,Weyland2004,Xu2015,Yang2017}, but requires a large number of images from different projections, which is difficult to obtain for radiation sensitive structures.

Here, we obtain the structure of an ultra-thin sample -- one atomic layer for the Si substitution, or up to two atoms behind each other in a projection for the Si tetramer -- from a small number of images with different sample tilts. We directly show the out-of-plane buckling induced by the Si dopant from the medium-angle annular dark-field (MAADF) STEM images with the sample (and hence the plane of the graphene sheet) tilted by approximately 20 degrees away from the normal incidence of the electron beam.  While the Si dopant in graphene appears symmetric in normal incidence plane-view images \cite{Ramasse2013}, the out-of-plane deformation of the atoms causes a significant symmetry breaking of the atomic positions in the tilted projection. First, we show that a computationally relaxed model of a silicon substitution in graphene matches significantly better to a tilted STEM image of this structure than to a flat one. This already indicates that the theoretical model agrees well with the experimental observation.
For a quantitative analysis of the 3D structure, we use an optimization process where an atomistic model is iteratively optimized to achieve the best possible fit to two experimental images with different viewing angles. Details of the method are described in Ref. \cite{Hofer2018b}. 

Since the electron beam can transfer more energy to Si in the graphene lattice than is required for flipping a buckled Si atom from one side of the lattice to the other \cite{Susi2014}, this process should be expected. As a consequence, one might expect all Si atoms to be one the side opposing the electron source. However our results reveal both configurations and the transition between the two.
We also show the 3D structure of a pyramid-like Si-tetramer, whose formation and observation was reported in Ref. \cite{Dyck}. Also in our case, the structure was formed from a Si trimer \cite{Yang2014} by the capture of an additional Si atom. Calculations show that the central Si atom in the tetrameric configuration is almost \SI{3}{\angstrom} above the graphene plane. The projected position of this atom is displaced relative to the other atoms when the sample is tilted. Also here, the 3D structure can be reconstructed by matching the model to the experimental images, and our reconstruction is in excellent agreement with the computer model. 
Similar to the single Si substitution, we observed a single flipping event presumably induced by the electron beam. Our results not only show the first experimental obtained 3D structures, but also insights into their electron beam induced out-of-plane dynamics.

\section{Results and Discussion}

\begin{figure}

\centering
\includegraphics[width=1\textwidth]{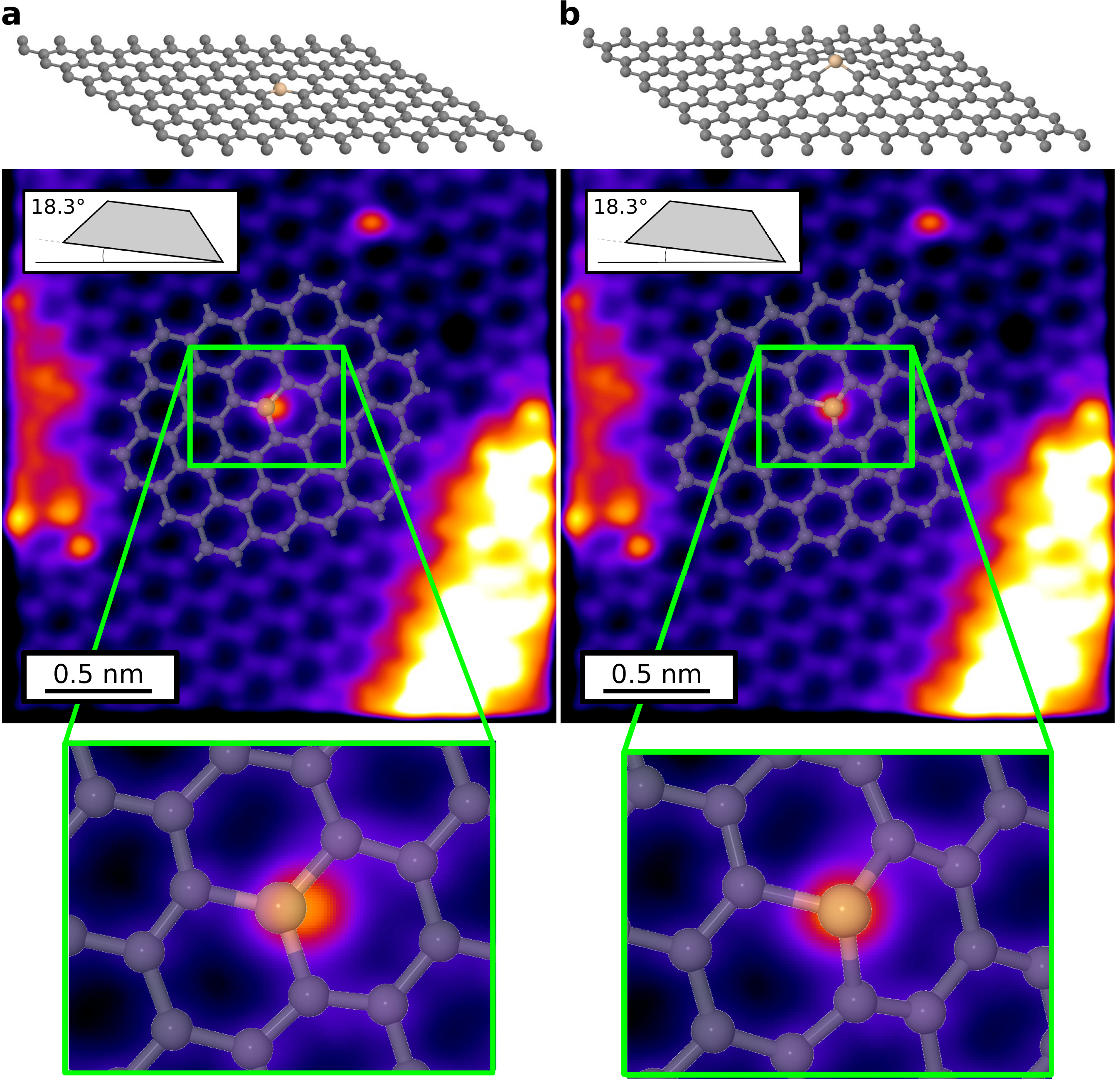}

\caption{ STEM image of a Si dopant in graphene overlaid by a flat model (a) and a relaxed model (b). }
\label{overlay}
\end{figure}

Figure \ref{overlay} shows a STEM image of a Si dopant with a sample tilt of ca. 18$^\circ$, overlayed by a tilted flat model (left panel) and a relaxed model (right panel -- see Methods). The model was scaled and translated so that the graphene lattice around the impurity matches the experimental image. In the flat structure, the expected Si position deviates from the experimentally observed position whereas the carbon positions match very well. However, a visually perfect match between the model and the STEM projection can be obtained by overlaying the computationally relaxed 3D model to the experimental image. This approach already qualitatively confirms that the theoretically predicted 3D structure can be also verified in the electron microscope.

\begin{figure}

\centering
\includegraphics[width=1\textwidth]{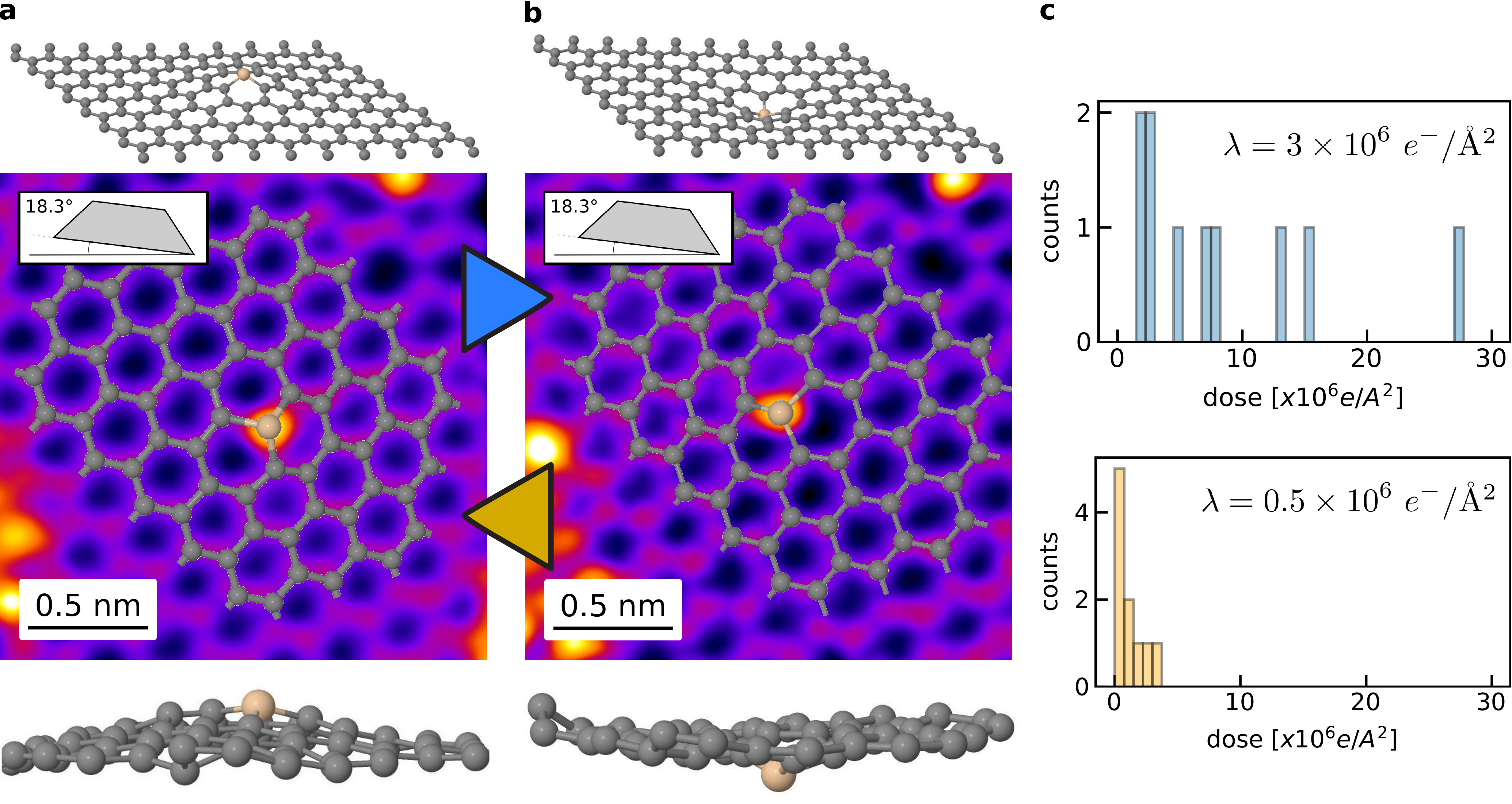}

\caption{STEM projections of a Si dopant in graphene with a sample tilt of ca. \SI{18}{\degree}. Panel a and b show that the experiment shows both "buckled up" and "buckled down" structures. The bottom row shows the actual 3D reconstruction. Panel c shows how much dose is required to flip the Si from below the graphene plane to above it and vice versa, respectively, in orange and blue. }
\label{Si}
\end{figure}

Figure \ref{Si} shows two different STEM images with a Si substitution with the same sample tilt as in Figure \ref{overlay}. From the tilt direction, the Si atom in panel (a) can be identified as sticking above the graphene plane (with respect to the electron beam coming from the bottom in our microscope). Although two energetically equivalent states of the silicon (buckling above and below the graphene plane) are expected, this configuration appears to be more frequent due to the direction of the electron beam. After several more scans, the same projected silicon atom is seen on the other side of the graphene plane (panel (b)). The stability of this configuration is, as expected to be, much lower than the previous one because the direction of the buckling is opposite to the direction of the momentum of the electron beam.
The expected value required to cause the dopant to flip from 'up' to 'down' and vice versa is calculated by a Poisson analysis. Here, the electron doses until a corresponding event occurs are collected and shown in the histograms in Figure \ref{Si}. In total, 10 events for each configuration are used for the analysis. The blue and the orange columns show the flip from above the graphene plane to below (towards the electron beam) and vice versa. A Poisson fit yields the mean values $\lambda_u = \SI{2.87e6}{\electrons\per\angstrom\squared} $ for the flip from up to down and $\lambda_d = \SI{4.96e5}{\electrons\per\angstrom\squared}$ for the flip from down to up.

Based on a DFT molecular dynamics simulation \cite{Susi2014}, the kinetic energy perpendicular to the graphene lattice that a probe electron would need to transfer to a Si substitution in order to shift it from one side of the graphene lattice to the other is between [1.375,1.50] eV. For our \SI{60}{keV} electron beam, this corresponds to a cross section of about 1100 barn \cite{Susi2016}. The experimentally observed areal event dose of \SI{0.5 e6}{\electrons\per\angstrom\squared } corresponds to a cross section value of 200 barn. On the other hand, the Si bouncing back to the side facing towards the electron beam was not observed in the molecular dynamics simulation. These discrepancies might be due to non-perpendicular momentum transfers due to sample tilt.

With experimental images of the "up" and "down" configuration at a fixed sample tilt, and a reference image of the impurity at zero sample tilt, we reconstruct the 3D configuration as described previously \cite{Hofer2018b}. In brief, an atomistic model is created, and the atoms are shifted in such a way that the difference between the experimental images and simulated images is minimized. The obtained 3D structures in the corresponding orientations are shown in the bottom row of Figure \ref{Si}. Besides the out-of-plane position of the Si impurity, they reveal slight buckling also of the surrounding graphene sheet. 
The Si-C bond length obtained from the experimental data via the 3D reconstructions is \SI[separate-uncertainty = true]{1.68(7)}{\angstrom}, which is close to the theoretical prediction (\SI{1.74}{\angstrom}).

\begin{figure}

\centering
\includegraphics[width=1\textwidth]{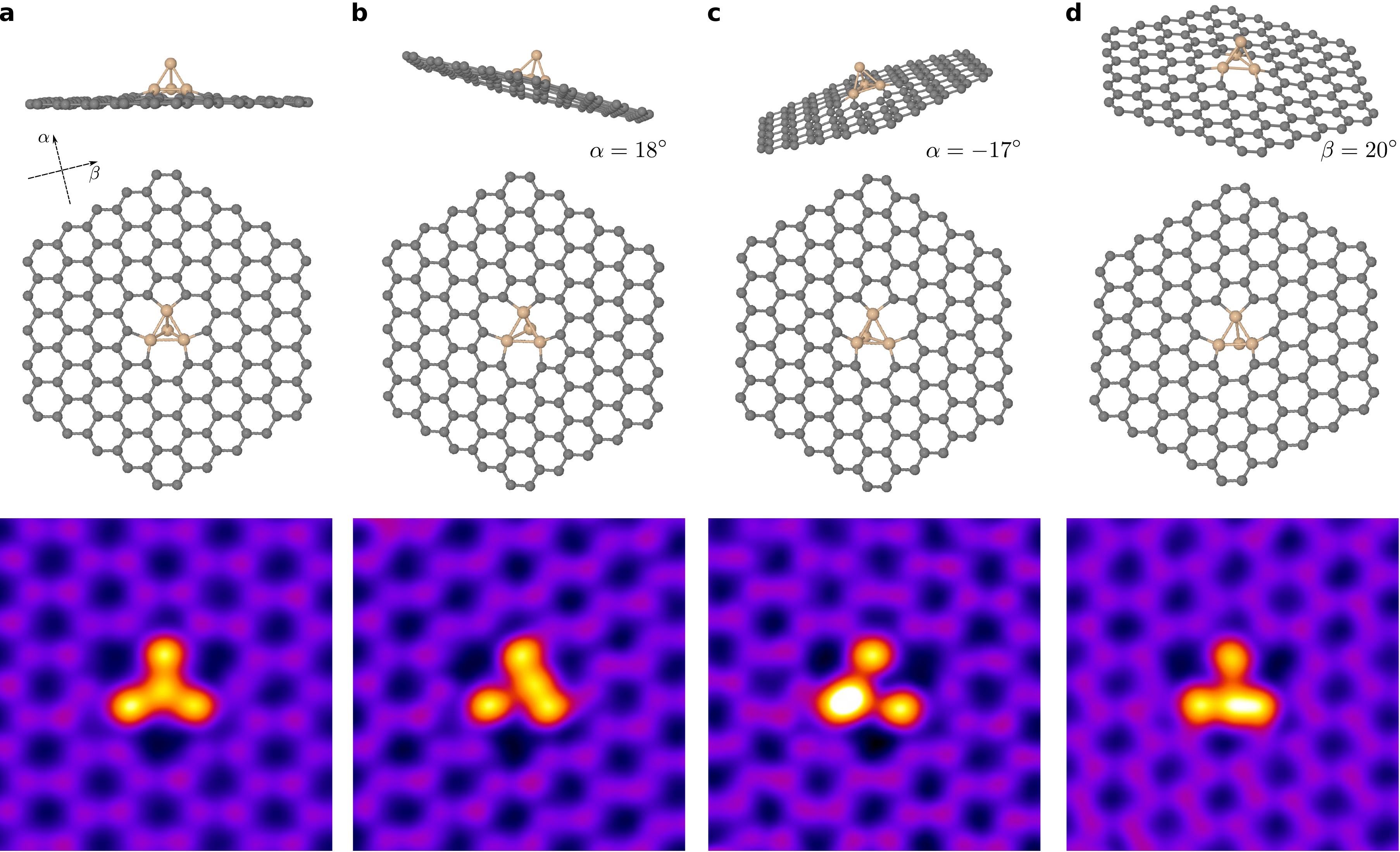}

\caption{Computational models and STEM projections of a Si tetramer in graphene with a non-tilted sample (a) and with a sample tilt of ca. \SI{20}{\degree} (b-d) in different directions. The highly asymmetric displacement of the central Si atom indicates a significantly out-of-plane position.}
\label{tetra}
\end{figure}

\begin{figure}

\centering
\includegraphics[width=1\textwidth]{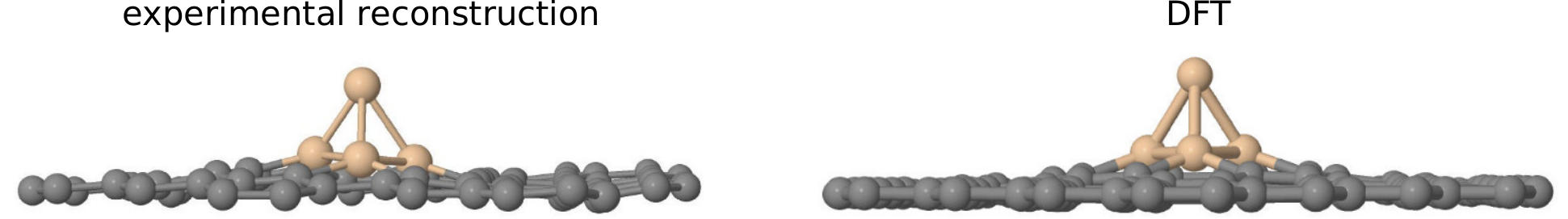}

\caption{3D structures of the Si trimer (a) and the Si tetramer (b) obtained by image reconstruction (left) and DFT calculation (right).}
\label{threeD}
\end{figure}

Next, we show the 3D structure of a Si tetramer. This structure is very beam stable allowing us to tilt the sample in different directions and reconstruct its three-dimensional shape. Computational studies of this structure reveal a triangular pyramid structure, where the central Si atom is sticking out of the plane by ca. \SI{3}{\angstrom}. The top row of Figure \ref{tetra} shows a side view of the relaxed model tilted in different directions. The top view (second row) shows the projected positions with respect to the electron beam. The experimental images (third row) agree well with the model. Again, we further reconstruct the 3D shape using all four projections, and the result matches precisely with calculations (Fig. \ref{threeD}).

\begin{figure}

\centering
\includegraphics[width=1\textwidth]{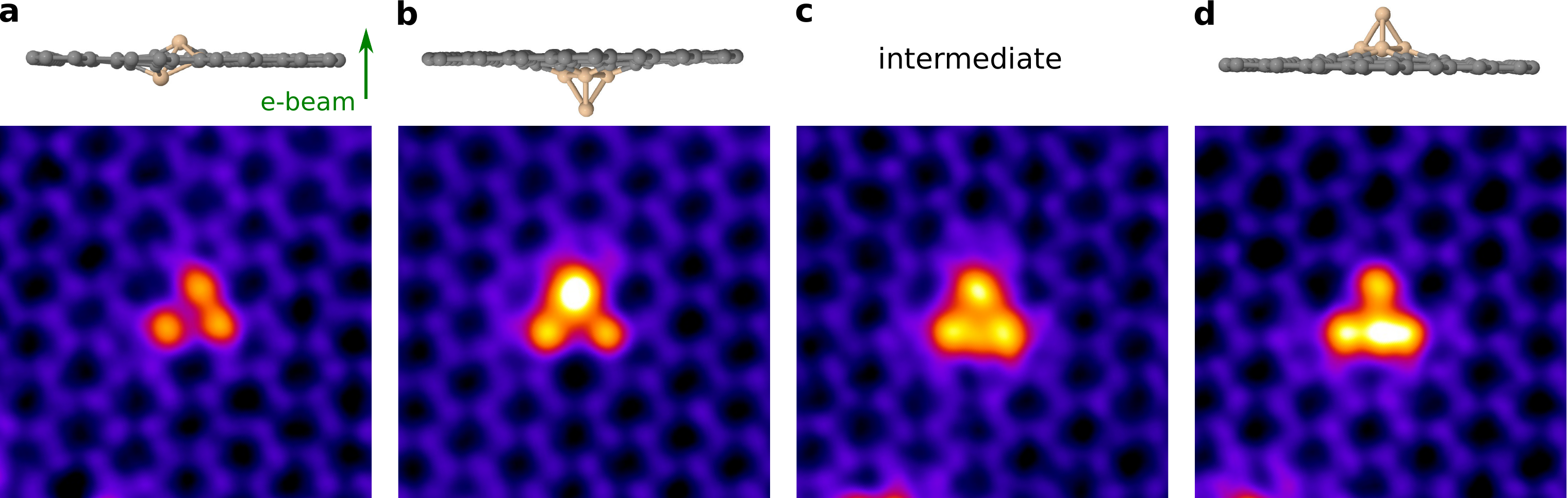}
\caption{Computational models and STEM projections of the Si tetramer in graphene with a sample tilt of ca. \SI{20}{\degree}. After a fourth Si atom formed the tetramer, the pyramid like structure transitioned after an electron dose of ca. \SI{1.6e7}{\electrons\per\square\angstrom} to the other side of graphene, facing away from the electron beam. This configuration did not change further. }
\label{tetrajump}

\end{figure}

We also observed a transition of the tetramer, where the Si\textsubscript{4} pyramid flipped from below to above the graphene plane. Also in this case, breaking the symmetry by tilting the sample is the key to observe the dynamics. Fig. \ref{tetrajump}a shows a STEM image of a Si trimer, which captures a fourth Si atom to create the tetramer (b). In this case, the Si was captured below the graphene plane. However, after some time the continuous electron irradiation triggers a transition to an intermediate step (panel c), which could not be clearly identified. Eventually, the Si atoms are arranged again as a tetramer, facing in the opposite direction with respect to the initial state. Hence, the whole tetramer was flipped to the opposite side of the graphene plane, where it remained stable for a large number of atomically resolved STEM images.

\section{Conclusions}

We have experimentally confirmed the out-of-plane structure of a single Si substitution and revealed the 3D structure of a Si tetramer in graphene. The single silicon dopant induced a buckling of the surrounding graphene sheet and we observed its electron beam induced flip in both directions, with an asymmetric cross-section. The Si tetramer is stable under the electron beam, after changing from opposing the electron beam to the other side of graphene. Our results provide new insights into the structure and dynamics of Si dopants in graphene, and the approach may be extended to other impurities and other 2D materials. 

\section*{Methods}

\subsection*{Sample}
For our experiment, we used graphene oxide (GO) which contains mobile C and Si adatoms as by-products. Water dispersion of GO was received from Danubia NanoTech, Ltd. The oxidation method of graphitic powder and subsequent exfoliation were developed by the company with an ultimate goal to preserve the long-range structural order in the graphene oxide flakes exfoliated down to the single-atom thickness. The product data-sheet is available at \url{https://www.danubiananotech.com/wp-content/uploads/Datasheet-GO_liquid-REV_21_12_2015.pdf}. Water dispersion of GO was significantly diluted ($\sim 1:100$ ); a TEM grid was then vertically dipped into the dispersion for a minute and dried in air afterwards.

\subsection*{Electron Microscopy}
Scanning transmission electron microscopy (STEM) experiments were conducted using a Nion UltraSTEM100, operated at \SI{60}{\kV}. Typically, our atomic-resolution images were recorded with $512\times 512$ pixels for a field of view of \numrange[range-phrase=--]{2}{4} nm  and dwell time of \SI{16}{\us} per pixel using the medium angle annular dark field (MAADF) detector. To enhance the signal, multiple identical images were averaged.

\subsection*{Density Functional Theory}
Density functional theory as implemented in the GPAW package was used to minimize structural models of the Si substitution \cite{Susi2014} and the Si\textsubscript{4} pyramid. For the latter case, a graphene supercell of $11 \times 11 \times 1$ containing 242 carbon atoms was constructed. Some of the carbon atoms were substituted with Si to construct a model matching the experimental observation. Employing the LCAO double zeta polarized basis set, a $\left| k \right|$-point mesh of $5 \times 5 \times 1$, the PBE functional and a computational grid spacing of \SI{0.2}{\angstrom}, the models were relaxed until the maximum forces were less than \SI{0.01}{\electronvolt\per\angstrom}. In addition, we estimated via molecular dynamics simulations \cite{Susi2014} the perpendicular kinetic energy required to shift a Si substitution from one side of the graphene lattice to the other.

\section*{Supporting Information}
In the supporting information, the whole set of raw data is provided.

\section*{Acknowledgments}

This work was supported by the European Research Council Starting Grant no. 336453-PICOMAT. M.R.A.M, and J.K. acknowledge support from the Austrian Science Fund (FWF) through project I3181-N26 and T.S . through project P28322-N36.

\end{document}